\documentclass[usenatbib,useAMS]{mnras}
\usepackage{epsfig}
\begin{document}
\title[Hamiltonian Formalism for dynamics of particles in MOG]{Hamiltonian Formalism for dynamics of particles in MOG}
\author[Sohrab Rahvar]
{Sohrab Rahvar$^{1}$\thanks{rahvar@sharif.edu}  \\
$1$ Department of Physics, Sharif University of Technology, P.O.
Box 11155-9161, Tehran, Iran }

\maketitle
\begin{abstract}
MOG as a modified gravity theory is designed to be replaced with dark matter. In this theory, in addition to the metric tensor, a massive vector is a gravity field where each particle has a charge proportional to the inertial mass and couples to the vector field through the four-velocity of a particle. 
In this work, we present the Hamiltonian formalism for the dynamics of particles in this theory. The advantage of Hamiltonian formalism is a better understanding and analyzing the dynamics of massive and massless particles. The massive particles deviate from the geodesics of space-time and photons follow the geodesics. We also study the dynamics of particles in the Newtonian and post-Newtonian regimes for observational purposes. An important result of Hamiltonian formalism is that while lensing on large scales is compatible with the observations,  however
the deflection angle from stellar size lensing is larger than General Relativity. This result can rule out this theory unless we introduce a screening mechanism to change the effective gravitational constant near compact objects like stars.\end{abstract}

\section{Introduction}
The dynamics of galaxies and large-scale structures of the Universe show that a significant amount of matter in the Universe is dark \citep{2005PhR...405..279B}. The observational evidence for the existence of dark matter started with the measurement of the rotation curve of spiral galaxies \citep{Rubin:1970zza}. We have a list of candidates for the dark matter, however, observations 
in recent years ruled out some of these candidates. For instance, the microlensing observations in the direction of Large and Small Magellanic clouds ruled out the Massive Astrophysical Compact Halo Objects (MACHOs) as the dark matter candidates \citep{2002SSRv..100..103M,2007A&A...469..387T,2011MNRAS.416.2949W}.

There are also experiments for the detection of non-baryonic candidates for the dark matter as axions, sterile neutrinos, weakly interacting massive particles (WIMPs) and supersymmetric particles \citep{2004PhR...402..267O} where until now no evidence for the dark matter particles is reported. The other candidate of dark matter is the Primordial Black Holes (PBHs) which could be formed in the early universe as a result of quantum fluctuations \citep{1966AZh....43..758Z,10.1093/mnras/152.1.75}. Various observations exclude them as the dark matter candidate except for the two narrow windows of lunar mass and tens of solar masses \citep{2020ARNPS..70..355C}. On the other hand, the PBHs might be the sources of large mass black holes ($m>50 m_\odot$) where in the gravitational wave events they have been detected by LIGO \citep{2021PhRvD.103h4001K}. Also, the lunar mass black holes may collide with the earth, however considering hundred percent of the halo is made of PBHs, their collision rate is one per billion year and has a weak signature on the earth \citep{2021MNRAS.507..914R}. So the PBHs could be a possible candidate for dark matter. 

In recent years, the lack of detection of dark matter candidates motivated the study of modification to the gravity law such as MOND \citep{1983ApJ...270..365M,2021PhRvL.127p1302S}. One of the modified theory models is so-called MOdified Gravity (MOG) \citep{Moffat:2005si} where in addition to metric as a gravity field there is also a vector and scalar sectors to the gravity. In this theory, each object in addition to the gravitational mass has a gravitational charge, proportional to the inertial mass that couples with a massive vector field. The result is that for the long distances from a point-like source of gravity the massive vector field fade and we recover the Einstein general relativity. In this case,  the gravitational constant (i.e. $G$) is tuned to be large at large scales to compensate for the dark matter and on the small scales the repulsive vector field weakens the gravitational strength and we will have an effective gravity with a smaller gravitational constant of $G_N$. 

The observational tests of this theory in the weak field approximation for the dynamics of galaxies \citep{Moffat:2013sja} and the cluster of galaxies \citep{Moffat:2013uaa} have been investigated. The comparison of data from observation with the prediction of MOG  shows that the dynamics of these structures can be interpreted without the need for dark matter even in the cosmological scales \citep{2020MNRAS.496.3502D,2021MNRAS.507.3387D}. One of the challenging 
problems to test the modified gravity models is the gravitational lensing from the large-scale structures. Since the 
mass of photons is zero, on the other hand, massive particles follow the modified geodesic world lines, it is confusing how to deal photons couples with the vector field. The wave optics approach of electromagnetic propagation has been used to study this problem \citep{2019MNRAS.482.4514R}. The observational comparison of the gravitational lensing on the large scales confirms that strong lensing by galaxies can be interpreted without the need for the dark matter \citep{2012arXiv1204.2985M}.

In this work, we introduce the Hamiltonian formalism for the dynamic of massive and massless particles in MOG. This approach resolves the ambiguity in the dynamics of particles, especially for the massless particles. We show that, unlike the massive particles, massless particles follow a different world line in this theory. In Section (\ref{fe}) we provide the action for this theory and the dynamics of particles in the Hamiltonian formalism. In Section (\ref{SE}) we introduce the field equation in MOG. In Section (\ref{WF}) we derive the equation of motion of particles in the weak field approximation and emphasis the gravitational lensing for the massless particles. In Section (\ref{pn}), we extend our calculation to the post-Newtonian limit and consider the perihelion precession of Mercury in MOG. Section (\ref{conc}) provides the conclusion.

\section{Field equation}
\label{fe}
In the simplified form of MOG theory, the gravity is given by the metric $g_{\mu\nu}$ and a vector field of $\phi_\mu$ where the overall action for the field equation can be written as \citep{Moffat:2005si}
 \begin{equation}
 \label{action}
S = S_g + S_\phi + S_M,
\end{equation}
 where the action associate to the metric is
 \begin{equation}
 S_g = \frac{c^4}{16\pi G}\int(R+2\Lambda) \sqrt{-g} d^4x, 
 \end{equation}
 and the action associated to the vector field is 
 \begin{eqnarray}
S_\phi&=&\frac{1}{4\pi}\int\omega\Big[-\frac{1}{4}B^{\mu\nu}B_{\mu\nu}+\frac{1}{2}\mu^2\phi_\mu\phi^\mu\nonumber\\
&+&  V_\phi(\phi_\mu\phi^\mu)\Big]\sqrt{-g}~d^4x - \int J_\mu \phi^\mu\sqrt{-g}~d^4x ,
\label{phi}
\end{eqnarray}
where we adapt $(-,+,+,+)$ signature for the metric, 
$\omega$ is coupling constant, $B_{\mu\nu}=\partial_\mu\phi_\nu-\partial_\nu\phi_\mu$ is Faraday tensor,  $J^\mu = \kappa \rho u^\mu$ and  $J_\mu\phi^\mu$ is the interaction term and $S_M$ is the matter action. The field equations are obtained by varying the action with respect to the metric and vector fields. 

On the other hand, for a test mass particle interacting with the metric and vector part of gravity the action is 
given by 
\begin{equation}
S = -m c^2\int d\tau +q_5 \omega \int \phi_\mu dx^\mu.
\end{equation}
The main assumption of this theory is that the charge of particle in the second term is proportional to the inertial mass of particle, i.e. $q_5 = \kappa m$. We write this action in terms of coordinate time in curved space as follows 
\begin{equation}
S = -mc^2\int (-\frac{1}{c^2}g_{\mu\nu} \frac{dx^\mu}{dt}\frac{dx^\nu}{dt})^{1/2} dt + m\kappa \omega \int \phi_\mu \frac{dx^\mu}{dt} dt .
\end{equation}
The Lagrangian corresponding to this action is 
\begin{equation}
\label{Lag}
L = -mc(-g_{ij} \dot{x}^i\dot{x}^j - 2g_{0i}c\dot{x}^i-c^2 g_{00})^{1/2}+m\kappa \omega (\phi_i\dot{x}^i+\phi_0 c), 
\end{equation}
where the latin indices corresponds to the spatial part of metric. The canonical momentum of particles from the definition of $p_i = \frac{\partial L}{\partial \dot{x}^i}$ obtain as  
\begin{equation}
\label{p}
p_i =\frac{ mc (g_{ij}\dot{x}^j + g_{0i}c)}{(-g_{ij} \dot{x}^i\dot{x}^j - 2g_{0i}c\dot{x}^i-c^2 g_{00})^{1/2}} + m\kappa\omega \phi_i. 
\end{equation}
According to the definition of Hamiltonian as $H = p_i \dot{x}^i - L$, the Hamiltonian of a particle is given by 
\begin{equation}
\label{H}
H = -mc^2\frac{g_{00}c + g_{0i}\dot{x}^i}{(-g_{ij} \dot{x}^i\dot{x}^j - 2g_{0i}c\dot{x}^i-c^2 g_{00})^{1/2}} - mc\kappa \omega \phi_0.
\end{equation}
We simplify this equation to write the Hamiltonian in a specific coordinate by setting $g_{0i} = 0$, then we can write the Hamiltonian as 
conventional form in terms of coordinate and momentum. Substituting equation (\ref{p}) in (\ref{H}), the Hamiltonian simplifies to 
\begin{equation}
\label{Ham0}
H = \sqrt{-g_{00}}E  - mc\kappa \omega \phi_0,
\end{equation}
where 
\begin{equation}
E = c\sqrt{(mc)^2 + p_i p^i + (\kappa \omega m)^2 \phi_i \phi^i - 2\kappa \omega m p_i \phi^i }. \nonumber
\end{equation}
 In what follows we obtain the dynamics of massive and massless particles in a generic space-time.
 From the Hamilton equations $\dot{p_k} = -\frac{\partial H}{\partial x_k}$ and $\dot{x}^k = \frac{\partial H}{\partial p_k}$, the dynamics of a test particle obtain as  
\begin{eqnarray}
\label{Ham1}
\dot{p_k} &=& - E (\sqrt{-g_{00}})_{,k} - \sqrt{-g_{00}} E_{,k} + mc\kappa \omega  \phi_{0,k} \\
\dot{x}^k &=& \frac{1}{E}{(p_j c^2- \kappa \omega mc^2 \phi_j)g^{jk}}.
\label{Ham2}
\end{eqnarray}
where 
\begin{equation}
E_{,k} = \frac{c}{E}\left(\frac12 p_ip_j g^{ij}{}_{,k} + (\kappa \omega m)^2\phi_i \phi^i{}_{,k} -\kappa \omega mp_i(\phi_{j,k} g^{ij} + \phi_j g^{ij}{}_{,k})\right)
\end{equation}
In the next section we introduce the field equations, then apply them to the Hamiltonian equations to obtain the dynamics of the particles. 

\section{Solution of field equations}
In this section, we review the field equations in MOG.
\label{SE}
 Varying action in equation (\ref{action}) with respect to the metric results in 
\begin{equation}
\label{G}
G_{\mu\nu} = \frac{8\pi G}{c^4} (T_{\mu\nu}^{(M)} + T_{\mu\nu}^{(\phi)}), 
\end{equation}
where $G_{\mu\nu} = R_{\mu\nu} - \frac12 R g_{\mu\nu} $ is the Einstein tensor and energy momentum tensor is defined by 
\begin{equation}
T_{\mu\nu}^{(M)} + T_{\mu\nu}^{(\phi)}= -\frac{2}{\sqrt{-g}}  \frac{\delta\left[(L_M+L_\phi)\sqrt{-g}\right]}{\delta g^{\mu\nu}},
\end{equation}
where the energy momentum tensor of vector field is 
\begin{equation}
T_{\mu\nu}^{(\phi)} = \frac{\omega}{4\pi}(B_\mu{}^\alpha B_{\nu\alpha} - \frac{1}{4}g_{\mu\nu}B^{\alpha\beta}B_{\alpha\beta})-\frac{\mu^2\omega}{4\pi}(\phi_\mu\phi_\nu - \frac12 g_{\mu\nu}\phi_\alpha \phi^\alpha).
\end{equation}
The first and second terms of the energy-momentum tensor are in the order of $\sim (\partial \phi)^2$ and $\sim \mu^2 \phi^2$. For the weak field approximation, the vector field and Newtonian gravitational potential are proportional as $|\phi |\propto \Phi_N/\sqrt{G}$. So  $\sim (\partial \phi)^2$ would be in the order of energy density of gravitational potential or $\sim \rho v^2$ \citep{Moffat:2013sja}. We can ignore this term compared to the matter rest mass energy density  (i.e. $T^{00} = \rho c^2$). 

So in the right hand side of modified gravity equation we ignore the energy momentum of vector field, $T_{\mu\nu}^{(\phi)}$ compare to the energy momentum of matter $T_{\mu\nu}^{(M)}$.

Now, we vary the action in equation (\ref{phi}) with respect to $\phi^\nu$. The result is 
\begin{equation}
\label{phis}
\phi^\nu{}_{;\mu}{}^{\mu} - \mu^2\phi^\nu = -\frac{4\pi}{\omega} J^\nu.
\end{equation}
In this work we study the static gravitational field where the time derivation of $\phi^\nu$ is zero. In cosmology where the all the fields are changing by time this argument is not valid and due to isotropy of the Universe, the spatial derivatives are zero. For static metric equation (\ref{phis}) simplifies to 
\begin{equation}
\label{phieq}
\phi^\nu{}_{,i}{}^{i} + \phi^\nu{}_{,}{}^{i}(\ln\sqrt{-g})_{,i} - \mu^2\phi^\nu = -\frac{4\pi}{\omega} J^\nu
\end{equation}
Now we discuss the solutions of equations (\ref{G}) and (\ref{phieq}) around the flat space.

\section{Dynamics in the weak field approximation}
\label{WF}
Let us assume the metric of weak field approximation up to the order of $g_{\mu\nu}\sim (v/c)^2$ as
\begin{equation}
\label{weakfield}
ds^2 = - (1+\frac{2\Phi}{c^2}) c^2dt^2 + (1-\frac{2\Phi}{c^2})\delta_{ij}dx^i dx^j.
\end{equation}
Here, the $2\Phi/c^2$ term is a perturbation around the flat space-time. Also, we assume $\phi_\mu$ as the vector sector of the gravity is also a perturbation around the flat space. In order to keep only the first order perturbations we ignore the higher order  terms contains $g^2$, $g\phi$ and $\phi^2$ in our calculations. Then the modified Einstein equation simplifies to the Poisson equation as 
\begin{equation}
\label{poisson}
\nabla^2\Phi = 4\pi G \rho
\end{equation}
 and equation (\ref{phieq}) simplifies to 
\begin{equation}
\label{phieq2}
\nabla^2\phi^\nu - \mu^2\phi^\nu =  -\frac{4\pi}{\omega} J^\nu.
\end{equation}
The solution of this differential equation is
\begin{equation}
\phi^\mu(x) = \frac{1}{\omega}\int \frac{e^{-\mu|{{\bf x}-{\bf x'}|}}}{|{\bf x}-{\bf x'}|} J^\mu({\bf x'})d^3x', 
\end{equation}
where for a distribution of matter we have non-zero source of current as $J^0 = \kappa \rho c$ and $J^i = \kappa \rho u^i$.
\subsection{Equation of motion for massive particles}
We substitute the metric components in the Hamiltonian equation of (\ref{Ham1}), the result is 
\begin{equation}
\label{pk}
\dot{p}_k = m(-\frac{\partial\Phi}{\partial x^{k}} + \kappa \omega \frac{p_i}{m} \frac{\partial\phi_{j}}{\partial x^k}\delta^{ij}+\kappa \omega c\frac{\partial \phi_{0}}{\partial x^k}).
\end{equation}
Here we assume that particles are moving with non-relativistic velocity (e.g $v/c\ll 1$). On the other hand 
the velocity of a particle from the Hamilton equation is 
\begin{equation}
\label{23}
\dot{x}^k = \delta^{ki}(\frac{p_i}{m} - \kappa\omega \phi_i),
\end{equation}
substituting the derivative of equation (\ref{23}) in (\ref{pk}), we obtain the dynamics of a test mass particle 
in the Newtonian approximation as follows:
\begin{equation}
\label{ddx}
\ddot{x}_k = -\frac{\partial}{\partial x^k}(\Phi + \kappa\omega c\phi^0) -\kappa\omega \frac{\partial \phi_k}{\partial t} + \kappa\omega [\vec{\dot{x}}\times(\nabla\times {\vec \phi})]_k,
\end{equation}
where we replace $\phi_0$  with the contravariant vector of $\phi^0$ in the first term.  We can define the effective potential of $\phi_{eff} = \Phi + \kappa\omega c \phi^0$ which is also derived in \cite{Moffat:2013sja}. 

In the Hamiltonian approach,  we derived the additional magneto-gravity terms in MOG. Let us define $-\kappa\omega \partial_t\phi_k$ as the "{\it Emog} term" where it is similar to an induced electric field due to time variation of the magnetic field. 
The third term on the right-hand side of this equation, $\kappa\omega \vec{v}\times(\nabla\times {\vec \phi})$ is similar to the Lorentz-force due to a magnetic term. Let us recall this term also "{\it Bmog} term". 
 
We note that in equation (\ref{ddx}) , we recover the Newtonian equation of $\ddot{x_k} = -\nabla\Phi_N$ plus the term of $~ \nabla \phi^0$ where the later term plays the role of dark matter in the large scale structures. These two terms are in the order of $v^2/r$. On the other hand, the {\it EMog} and {\it BMog} terms are in the order of ($\frac{v^4}{rc^2}$) which is in the post-Newtonian correction terms. We will discuss the contribution of these terms to the dynamics of a test particle around small structures such as stars and large structures such as galaxies. 

\subsubsection{The case for spherical symmetry}
Let us assume a point mass object like a star as the source of gravity. From the Poisson equation (\ref{poisson}) and equation (\ref{phieq2}), in the Newtonian regime the potentials are
\begin{eqnarray}
\Phi(r) &=& -\frac{GM}{r}, \\
\phi^0(r) &=& \frac{\kappa c}{\omega}\frac{Me^{-\mu r}}{r}
\end{eqnarray}
where $M$ is the mass of central star and $r$ is the distance from the star. The poisson term is an attractive force while the second term produces a repulsive force. For a star with relative velocity to the coordinate system we may also consider the magnetic component $\phi^i$. For the case of $J^i=0$ in equation (\ref{phieq2}), for the spherically symmetry we  have the following solution of
\begin{equation}
\label{mono}
\phi^r(r) = \frac{1}{\omega}\frac{c g e^{-\mu r}}{r} \hat{e}_r,
\end{equation}
where $g$ is a charge for the magnetic monopole and we can take it to be proportional to the inertial mass as $g = \kappa M$.  The solution for the dynamics of a point mass particle in equation (\ref{ddx}) up to $(v/c)^2$ terms and ignoring the 
post-Newtonian orders results in an attractive force as $1/r^2$ and a Yukawa type repulsive force.  The repulsive force at  large distances fades to zero and at the distance of $r<\mu^{-1}$ weakens the effective gravity.  Now we substitute the potentials in the equation of motion of a test particle in equation (\ref{ddx}),   
\begin{equation}
\label{mm}
\ddot{x}^i = -\frac{GM}{r}\left(1 - \frac{\kappa^2 c^2}{G} e^{-\mu r}(1+\mu r)\right).
\end{equation}

In equation (\ref{mm}) at the large distances the exponential terms goes to zero and dynamics of a test particle converge to $\ddot{x}^i = -GM/r$. On the other hand, for the closer distances $r\ll \mu^{-1}$, the dynamics of a test particle is given by $\ddot{x}^i = -(G-\kappa^2 c^2)M/r$. Since for the closer distance we should recover the Newtonian gravity then $G_N = G - \kappa^2 c^2$. This means that the gravitational constant of theory (e.g. $G$) is larger than $G_N$ (defined at the smaller scales). Using the convention of $\kappa^2 c^2= \alpha G_N$, we can rewrite $G$ as $G = G_N(1+\alpha)$ where the best fit to the spiral galaxies results in $\alpha = 8.89 \pm 0.34$  \citep{Moffat:2013sja}. For the extended spherical structure as an elliptical galaxy we can obtain the effective potential of this structure as
\begin{equation}
\label{master}
\phi_{eff} = -G_N\int\frac{\rho(x')}{|x-x'|}(1+\alpha-\alpha e^{-\mu|x-x'|})d^3x'
\end{equation}

\subsection{Equation of motion of massless particles}
For the dynamics of massless particles as photons and neutrinos we start with the Hamiltonian in equation (\ref{Ham0}) and let $m =0$,  
\begin{equation}
H = \sqrt{-g_{00}}(p_ip_jg^{ij}c^2)^{1/2}.
\end{equation} 
From the Hamiltonian equations in (\ref{Ham1}) and (\ref{Ham2}), the time derivation of momentum and position obtain as
\begin{eqnarray}
\label{L1}
\frac{\dot{p}_k}{|p|} &=& - \frac{\partial\sqrt{-g_{00}}}{\partial x^k}c
-\frac{1}{2}\sqrt{-g_{00}}\frac{p_ip_j }{p^2}\frac{\partial g^{ij}}{\partial x^k}c, \\
\dot{x}^k &=& \frac{p_j}{p}g^{jk} \sqrt{-g_{00}}c.
\label{L2}
\end{eqnarray}
The main difference between the massive and massless particles is that the unlike to the massive particles that interact with $\phi_\mu$ field and deviate from the geodesics of the space-time, massless particles decouple from the vector field and follows the geodesics of metric. We substitute the weak field approximation of metric from 
equation (\ref{weakfield}) in equations (\ref{L1}) and (\ref{L2})  which results in the equation of motion of 
\begin{eqnarray}
\label{L3}
\frac{\dot{p}_k}{|p|} &=& -\frac{2}{c} \frac{\partial \Phi}{\partial x^k}\\
\dot{x}^k &=& \hat{n}^k (1+\frac{\Phi}{c^2}) c,
\label{L4}
\end{eqnarray}
where $\hat{n}^k=p^k/|p|$ is the unit vector along the light ray. 

Equation (\ref{L3}) represents the deflection of light while passing close to a gravitational potential. The change in the transverse component of the momentum of photon to the initial momentum represents the deflection angle of light 
where integrating along a light ray passing close to a mass with the impact factor of $b$ results in the deflection angle of 
$\delta = \frac{4\Phi(b)}{c^2}$.  For a single lens the deflection angle would be 
\begin{equation}
\delta_{MOG} = \frac{4G_NM(1+\alpha)}{bc^2}
\end{equation}
If we compare this deflection angle with the deflection angle from the Einstein gravity (i.e. $\alpha_{GR}$), we have enhancement in the deflection angle with the amount of $\delta_{MOG} = (1+\alpha) \delta_{GR}$. We note that this result is independent of the mass of the lens and impact factor of the light. One of the solutions to solve this problem might be the screening mechanism where close to the compact object the effective $G$ becomes smaller. The screening mechanism is practical for the scalar fields \citep{2010arXiv1011.5909K} and one may take $G$ as in the Brans-Dick theory to hide the enhancement of $G$ in the solar system scales. 

For a spherical symmetry-static space-time, taking into $G$ as the scalar field in this theory, the solution of the field equation results in a distance dependent value for the $\alpha$ parameter as \citep{2009CQGra..26h5002M}
\begin{equation}
\alpha = \frac{M}{(\sqrt{M} + {\cal E})^2}(\frac{G}{G_N} - 1),
\end{equation}
where ${\cal E}^2\gg M_\odot$. So for the stellar mass object the $\alpha$ parameter would be zero and we recover the gravitational lensing as in General Relativity.

\section{MOG Dynamics in post-Newtonian approximation}
\label{pn}
In this section we extend the approximation for the dynamics of particles (e.g $v/c$) in MOG to the higher orders and investigate the observational effects in the astronomical systems.

In the standard formalism of general relativity, one of the methods to investigate the relativistic effects in the slow-moving objects as the planets in the solar system is the post-Newtonian approximation \citep{weinberg}. In the Newtonian approximation of GR the equation of motion is given by $\dot{v^i} = \frac12 c^2 \partial^i h_{00}$ where the perturbation of metric is in the order of $\phi/c^2$ or $(v/c)^2$.  According to the convention, we call the perturbation of the metric by $g^{(N)}_{\mu\nu}$ where $N$ represents its order of magnitude in $(v/c)^N$. So the Newtonian approximation is the perturbation theory up to $N=2$. From equation (\ref{Lag}), the Lagrangian of particle for Post-Newtonian term would be 
\begin{equation} 
L = mc^2(\frac12 \bar{v}^2+\frac12 g_{00}^{(2)}+\frac12g_{00}^{(4)}+g_{0i}^{(3)}\bar{v}^i+\frac12 g^{(2)}_{ij}\bar{v}^i\bar{v}^j )+mc\kappa \omega(\phi_i\bar{v}^i+\phi_0), 
\end{equation}
where $\bar{v}^i$ is the physical velocity normalized to the speed of light. We have seen in the previous section that the "Bmog" and "Emog" terms also are in the order of $\bar{v}^4$, so these two terms can be considered as the post-Newtonian terms. Using the Lagrangian equation
\begin{equation}
\frac{d}{dt}\frac{\partial L}{\partial v^i} - \frac{\partial L}{\partial x^i} = 0, 
\end{equation}
the equation of motion of a test particle obtain as 
\begin{eqnarray}
\frac{d{\bf v}}{dt} &=& -{\bf \nabla}\phi_{eff} -\kappa\omega \frac{\partial {\vec \phi}}{\partial t} + \kappa\omega [{\bf v}\times(\nabla\times {\vec \phi})] - \nabla(2\Phi^2 + \psi) \nonumber \\
&-&\frac1c \frac{\partial {\bf \xi}}{\partial t}+\frac{\bf v}{c}\times (\nabla \times {\bf \xi}) + 3\frac{\bf  v}{c^2}\frac{\partial \Phi}{\partial t} + 4\frac{{\bf v}}{c}(\frac{\bf v}{c}  \cdot\nabla)\Phi - (\frac{v}{c})^2\nabla\Phi, \nonumber \\
&&
\end{eqnarray}
where $\phi_{eff} = \Phi + \kappa\omega c\phi^0$, the second and third terms at the right hand side are Emog and Bmog terms, $g_{00}^{(2)} = -2\Phi/c^2$, $g_{ij}^{(2)} = -2\delta_{ij}\Phi/c^2$, $g_{0i}^{(3)} = \xi_i$ and $g_{00}^{(4)} = -2(\Phi/c^2)^2 - 2\psi$. 

For the static space-time, we can ignore the time derivatives of the metric components. Also for the {\it Bmog} term since the current is zero (i.e. $J^i$) then for the monopole solution of equation (\ref{phieq2}) $\vec{\phi} \sim \frac{e^{\mu r}}{r} \hat{e}_r$ where $\nabla\times \vec{\phi} = 0$. Then we expect that for the precession of Mercury's perihelion only the post-Newtonian standard terms in GR matter and MOG has no contribution to this effect. In another word, MOG is compatible with the precession measurement of Mercury.

The effect of Emog and BMog terms in the galactic dynamics compare to the dominant term of $-\nabla \phi_{eff}$ is smaller by the factor of $(v/c)^2$. Assuming the velocity of stars inside the galaxy in the order of $v = 200$ km/s, we expect that post-Newtonian, as well as Emog and BMog terms, would be six orders of magnitude smaller than the effective potential. Since the observational accuracy is not high enough we may ignore the contribution of these terms in studying galactic dynamics.

\section{Conclusion}
\label{conc}
In this work, we present the Hamiltonian formalism for the dynamics of particles in the MOG theory. The advantage of 
using the Hamiltonian formalism is that we can investigate the dynamics of the spatial coordinate of particles in terms of the physical time both for the massive and massless particles. For the massive particles in addition to the conventional terms in  \cite{Moffat:2013sja}, we derived the Emog and Bmog terms to represent the time variation of a vector field and Lorentz force in the dynamics of particles. These two terms are in the order of post-Newtonian correction. 

We also derived the equation of motion of massless particles in this theory where it is shown that massless particles unlike massive particles do not couple to the vector field. So photons follow the geodesics equation that is given by the metric of space-time. Hence taking into account the gravitational constant of theory ($G$) which is almost one order of magnitude larger than the Newtonian constant of $G_N$, the deflection angle for the large scale structures as the galaxies and clusters of galaxies provides a stronger light deflection which can be interpreted as the dark matter. 
For the stellar mass lenses, this theory provides a stronger deflection angle unless $G$  is taken into account as an extra field in this theory. 

We also obtained the post-Newtonian approximation of the equation of motion in MOG. For a point mass object like the sun in the solar system, the Emog and Bmog terms are zero and only the standard post-Newtonian terms contribute to the precession of Mercury's perihelion. Also in the Galactic scales, while Emog and Bmog are non-zero, we can ignore the post-Newtonian terms as the accuracy of the observations is lower than the contribution of these terms.

\section*{Acknowledgments}
I would like to thank Shant Baghram for 
his useful comments.  Also, I would like to thank anonymous referee to his/her comments improving this work.
 This research was supported by Sharif University of Technology’s Office of Vice 
President for Research under Grant No. G950214. 
 
{\bf Data Availability:}
 No new data were generated or analysed in support of this research.

\bibliographystyle{mnras}
 	\bibliography{ref}

\end{document}